\documentstyle[preprint,aps]{revtex}

\begin{document}

\draft
\title
{Fractal Von Neumann Entropy}

\author
{Wellington da Cruz\footnote{E-mail: wdacruz@exatas.uel.br}}

\address
{Departamento de F\'{\i}sica,\\
 Universidade Estadual de Londrina, Caixa Postal 6001,\\
Cep 86051-970 Londrina, PR, Brazil\\}
 
\date{\today}

\maketitle

\begin{abstract}

We consider the {\it fractal von Neumann entropy}  associated with 
the {\it fractal distribution function} and we obtain for some {\it universal 
classes h of fractons } their entropies. We obtain also for each of 
these classes a {\it fractal-deformed Heisenberg algebra}. This one takes 
into account the braid group structure 
of these objects which live in two-dimensional multiply connected space.
  
\end{abstract}


The {\it fractal von Neumann entropy} associated with the 
{\it fractal distribution function } was introduced in Ref.\cite{R1}. 
We have defined {\it universal classes $h$ of particles or 
quasiparticles} which carry rational or irrational spin quantum number $s$.
The fractal parameter or Hausdorff 
dimension $h$ runs within the interval $1$$\;$$ < $$\;$$h$$\;$$ <$$\;$$ 2$. 
{\it Fractons} as charge-flux systems are such particles which live in 
two-dimensional multiply connected space and satisfy specific fractal distribution 
function. The particles of each class are collected taking 
into account the {\it fractal spectrum}

\begin{eqnarray}
&&h-1=1-\nu,\;\;\;\; 0 < \nu < 1;\;\;\;\;\;\;\;\;
 h-1=\nu-1,\;
\;\;\;\;\;\; 1 <\nu < 2;\\
&&etc.\nonumber
\end{eqnarray}

\noindent and the spin-statistics relation $\nu=2s$, 
which is valid for such fractons . For example, 
consider the universal classes with distinct values of spin $
\biggl\{\frac{1}{2},\frac{3}{2},\frac{5}{2},\cdots\biggr\}_{h=1}$, $
\biggl\{0,1,2,\cdots\biggr\}_{h=2}$ and $
\biggl\{\frac{1}{4},\frac{3}{4},\frac{5}{4},\cdots\biggr\}_{h=\frac{3}{2}}$, 
then we have the Fermi-Dirac distribution

\begin{eqnarray}
n[1]=\frac{1}{\xi+1}
\end{eqnarray}

\noindent the Bose-Einstein distribution

\begin{eqnarray}
n[2]=\frac{1}{\xi-1}
\end{eqnarray}

\noindent and the Fractal  distribution

\begin{eqnarray}
n\left[\frac{3}{2}\right]=\frac{1}{\sqrt{\frac{1}{4}+\xi^2}}
\end{eqnarray}

\noindent with $\xi=\exp\left\{(\epsilon-\mu)/KT\right\}$.

Another result of our approach to fractional spin 
particles is that the classes 
$h$ satisfy 
a {\it duality simmetry} defined by ${\tilde{h}}=3-h$, and as a 
consequence we extract a 
{\it fractal supersymmetry} which defines pairs of particles $\left(s,s+
\frac{1}{2}\right)$. Thus, {\it the fractal distribution function 
is a} 
{\bf quantum-geometrical} {\it description of the statistical laws of Nature, 
since the quantum path is a fractal curve and this reflects the 
Heisenberg uncertainty principle}.

The {\bf Fractal von Neumann entropy} per state in terms 
of the average occupation number is given as\cite{R1}

\begin{eqnarray}
\label{e5}
{\cal{S}}_{G}[h]&=& n\;K\biggl[({\cal{Y}}[\xi]-1)\ln({\cal{Y}}[\xi]-1)-
({\cal{Y}}[\xi]-2)\ln({\cal{Y}}[\xi]-2)\biggr]\\
\nonumber\\
&=& K\left[\left[1+(h-1)n\right]\ln\left\{\frac{1+(h-1)n}{n}\right\}
-\left[1+(h-2)n\right]\ln\left\{\frac{1+(h-2)n}{n}\right\}\right]\\
\nonumber
\end{eqnarray}

\noindent where 

\begin{eqnarray}
\label{e.h} 
n=\frac{1}{{\cal{Y}}[\xi]-h}
\end{eqnarray}

\noindent is the {\bf Fractal distribution function} {\it which appears 
as a simple and elegant generalization of the fermionic  
and bosonic distributions for particles with 
braiding properties}. The function 
${\cal{Y}}[\xi]$ satisfies the equation 

\begin{eqnarray}
\xi=\biggl\{{\cal{Y}}[\xi]-1\biggr\}^{h-1}
\biggl\{{\cal{Y}}[\xi]-2\biggr\}^{2-h}.
\end{eqnarray}

\noindent We also have 
 
\begin{eqnarray}
\xi^{-1}=\biggl\{\Theta[{\cal{Y}}]\biggr\}^{h-2}-
\biggl\{\Theta[{\cal{Y}}]\biggr\}^{h-1}
\end{eqnarray}

\noindent where

\begin{eqnarray}
\Theta[{\cal{Y}}]=
\frac{{\cal{Y}}[\xi]-2}{{\cal{Y}}[\xi]-1}
\end{eqnarray}

\noindent is the {\it single-particle 
partition function}.  The free energy for the particles in a given 
quantum state is expressed as 

\begin{eqnarray}
{\cal{F}}[h]=KT\ln\Theta[{\cal{Y}}],
\end{eqnarray}

\noindent  such that for fermions 

\begin{eqnarray}
{\cal{F}}[1]=-KT\ln\left\{1+\xi^{-1}\right\}
\end{eqnarray}

\noindent and for bosons 

\begin{eqnarray}
{\cal{F}}[2]=KT\ln\left\{1-\xi^{-1}\right\}.
\end{eqnarray}

\noindent Hence, we find the average occupation number

\begin{eqnarray}
\label{e.h} 
n[h]&=&\xi\frac{\partial}{\partial{\xi}}\ln\Theta[{\cal{Y}}]
=\frac{1}{KT}\xi\frac{\partial\;{\cal{F}}}{\partial{\xi}}.
\end{eqnarray}

\noindent Using the thermodynamic relation ${\cal{S}}=-
\biggl(\frac{\partial\;{\cal{F}}}{\partial{T}}\biggr)_{V,N}$, we obtain another 
general expression for the entropy

\begin{eqnarray}
\label{e.0} 
{\cal{S}}_{G}[h]&=&-K\ln\Theta[{\cal{Y}}]
-\frac{KT}{\Theta[{\cal{Y}}]}\;
\frac{\partial\;\Theta[{\cal{Y}}]}{\partial{T}}\\
\nonumber\\
&=&-K\ln\Theta[{\cal{Y}}]-\frac{KT}
{({\cal{Y}}-2)({\cal{Y}}-1)}\frac{\partial\;{\cal{Y}}}
{\partial{T}},
\end{eqnarray}

\noindent where the energy is given by

\begin{eqnarray}
{\cal{E}}&=&-\frac{KT^2}{\Theta[{\cal{Y}}]}\;
\frac{\partial\;\Theta[{\cal{Y}}]}{\partial{T}}\\
\nonumber\\
&=&-\frac{KT^2}{({\cal{Y}}-2)({\cal{Y}}-1)}
\frac{\partial\;{\cal{Y}}}{\partial{T}},
\end{eqnarray}

\noindent since we have ${\cal{E}}={\cal{F}}+T{\cal{S}}$. 

Now, if you consider the density matrix $\rho$, the entropy 
is found to be\cite{R2} 

\begin{eqnarray}
\label{e10} 
\frac{{\cal S}}{K}&=&- {\rm Tr}\;\rho\ln\rho\\
&=&-\sum_{n}{\cal W}(n){\cal P}(n)\ln{\cal P}(n)\\
\nonumber
\end{eqnarray}

\noindent where 
 
\begin{equation}
\label{e11}
{\cal W}(n)=\frac{\left[G+(nG-1)(h-1)\right]!}{[nG]!
\left[G+(nG-1)(h-1)-nG\right]!}
\end{equation}

\noindent is the statistical weight for the classes of fractons 
and $N=nG$ is the number of particles, $G$ is the number of states, 
$n$ is the average occupation number\cite{R1}. The microstate 
probability is given by

\begin{equation}
\label{e12}
{\cal P}(n)=p^{nG}q^{[n(h-2)+1]G},
\end{equation}

\noindent with $p+q=1$ and the total probability is unity 

\begin{equation}
\sum_{n}{\cal W}(n){\cal P}(n)=1.
\end{equation}

\noindent Differentiating this last expression with respect to $p$, we find 

\begin{equation}
n\left[\frac{q}{p}+2\right]-nh=1
\end{equation}

\noindent and defining $\left[\frac{q}{p}+2\right]\equiv\cal{Y}[\xi]$, 
we obtain again the fractal distribution function. Considering now the 
Eq.(20) plus the Eqs.(\ref{e11}) and (\ref{e12}) we have another 
derivation of the Eqs.(\ref{e5}) and (6) for the fractal von Neumann entropy.

The entropies for fermions $\biggl\{\frac{1}{2},
\frac{3}{2},\frac{5}{2},\cdots\biggr\}_{h=1}$ 
and bosons $\biggl\{0,1,2,\cdots\biggr\}_{h=2}$, 
can be recovered promptly
 
\begin{eqnarray}
{\cal{S}}_{G}[1]=-K\biggl\{n\ln n +(1-n)\ln (1-n)\biggr\} 
\end{eqnarray}

\noindent and
 
\begin{eqnarray}
{\cal{S}}_{G}[2]=K\biggl\{(1+n)\ln (1+n)-n\ln n\biggr\}. 
\end{eqnarray}

\noindent For fractons of the self-dual class $\biggl\{\frac{1}{4},
\frac{3}{4},\frac{5}{4},\cdots\biggr\}_{h=\frac{3}{2}}$, we have

\begin{eqnarray}
{\cal{S}}_{G}\left[\frac{3}{2}\right]=K\left\{(2+n)\ln\sqrt{\frac{2+n}{2n}}
-(2-n)\ln\sqrt{\frac{2-n}{2n}}\right\} 
\end{eqnarray}

\noindent and for two more examples, the dual classes $\biggl\{\frac{1}{3},
\frac{2}{3},\frac{4}{3},\cdots\biggr\}_{h=\frac{4}{3}}$ and 
$\biggl\{\frac{1}{6},\frac{5}{6},\frac{7}{6},\cdots\biggr\}_{h=\frac{5}{3}}$, \\
the entropies read as

\begin{eqnarray}
{\cal{S}}_{G}\left[\frac{4}{3}\right]=K\left\{(3+n)\ln\sqrt[3]{\frac{3+n}{3n}}
-(3-2n)\ln\sqrt[3]{\frac{3-2n}{3n}}\right\} 
\end{eqnarray}

\noindent and

\begin{eqnarray}
{\cal{S}}_{G}\left[\frac{5}{3}\right]=K\left\{(3+2n)\ln\sqrt[3]{\frac{3+2n}{3n}}
-(3-n)\ln\sqrt[3]{\frac{3-n}{3n}}\right\}. 
\end{eqnarray}

We have also introduced the topological concept of {\it fractal index} 
asssociated with each class ( $h$ is a geometrical parameter related 
to the quantum path of the particles ) and defined by\cite{R3}

\begin{equation}
\label{e.1}
i_{f}[h]=\frac{6}{\pi^2}\int_{\infty(T=0)}^{1(T=\infty)}
\frac{d\xi}{\xi}\ln\left\{\Theta[\cal{Y}(\xi)]\right\}
\end{equation}

\noindent so we obtain for the bosonic class $i_{f}[2]=1$, for the 
fermionic class $i_{f}[1]=0.5$ and for some classes of fractons  we have 
$i_{f}[\frac{3}{2}]=0.6$, $i_{f}[\frac{4}{3}]=0.56$, 
$i_{f}[\frac{5}{3}]=0.656$ . Thus for 
the interval of definition  $ 1$$\;$$ \leq $$\;$$h$$\;$$ \leq $$\;$$ 2$, 
there exists the correspondence $0.5$$\;$$ 
\leq $$\;$$i_{f}[h]$$\;$$ \leq $$\;$$ 1$, which signalizes the connection 
between fractons and quasiparticles of the conformal field theories, 
in accordance 
with the unitary
$c$$\;$$ <$$\;$$ 1$ representations of the central 
charge.

\noindent On the other hand, fractons satisfy a 
{\it fractal-deformed Heisenberg algebra} 
which generalizes the fermionic and bosonic ones. 
The fractal-deformed Heisenberg 
algebra is obtained of the relation

\begin{eqnarray}
{\bf a}(x){\bf a}^{\dagger}(y)-f[\pm h]{\bf a}^{\dagger}(y)
{\bf a}(x)=\delta (x-y),
\end{eqnarray}

\noindent between creation and annihilation operators. The factor of 
deformation is defined as 

\begin{eqnarray}
f[\pm h]=exp\left(\pm\imath h\pi\right), 
\end{eqnarray}

\noindent such that for $h=1$ and $x=y$, we reobtain the fermionic 
anticommutation relations $\left\{{\bf a}(x),
{\bf a}^{\dagger}(x)\right\}=1$, and for  $h=2$ and $x=y$, we 
reobtain the bosonic 
commutation relations $\left[{\bf a}(x),
{\bf a}^{\dagger}(x)\right]=1$. If $x\neq y$ and 
$1$$\;$$ < $$\;$$h$$\;$$ <$$\;$$ 2$, we have nonlocal operators 
for fractons

\begin{eqnarray}
{\bf a}(x){\bf a}^{\dagger}(y)=f[\pm h]{\bf a}^{\dagger}(y)
{\bf a}(x).
\end{eqnarray}

\noindent The braiding relations have the hermiticities

\begin{eqnarray}
{\bf a}(x){\bf a}^{\dagger}(y)&=&f[+h]{\bf a}^{\dagger}(y){\bf a}(x)\\
{\bf a}^{\dagger}(x){\bf a}(y)&=&f[+h]{\bf a}(y){\bf a}^{\dagger}(x)\\
\nonumber\\
{\bf a}(x){\bf a}^{\dagger}(y)&=&f[-h]{\bf a}^{\dagger}(y){\bf a}(x)\\
{\bf a}^{\dagger}(x){\bf a}(y)&=&f[-h]{\bf a}(y){\bf a}^{\dagger}(x)\\
\nonumber\\
{\bf a}(x){\bf a}(y)&=&f[+h]{\bf a}(y){\bf a}(x)\\
{\bf a}^{\dagger}(x){\bf a}^{\dagger}(y)&=&f[+h]{\bf a}^{\dagger}(y)
{\bf a}^{\dagger}(x)\\
\nonumber\\
{\bf a}(x){\bf a}(y)&=&f[-h]{\bf a}(y){\bf a}(x)\\
{\bf a}^{\dagger}(x){\bf a}^{\dagger}(y)&=&f[-h]{\bf a}^{\dagger}(y)
{\bf a}^{\dagger}(x),
\end{eqnarray}

\noindent where the plus sign of $h$ stands for anticlockwise exchange and 
the minus sign of $h$ for clockwise exchange.

\noindent The statistics parameter $\nu$ for the plus sign has the pattern 

\begin{eqnarray}
\left\{-,+,-,+,\cdots\right\}_{h}
\end{eqnarray}

\noindent and the minus sign, gives us another one

\begin{eqnarray}
\left\{+,-,+,-,\cdots\right\}_{h}.
\end{eqnarray}

\noindent For example, consider the class $h=\frac{3}{2}$, then

\begin{eqnarray}
\biggl\{\mp\frac{1}{2},\pm\frac{3}{2},\mp\frac{5}{2},
\pm\frac{7}{2},\cdots\biggr\}_{h=\frac{3}{2}},
\end{eqnarray}

\noindent i.e. 

\begin{eqnarray}
f[+h]=e^{-\imath \frac{1}{2}\pi}=e^{+\imath \frac{3}{2}\pi}
=e^{-\imath \frac{5}{2}\pi}=e^{+\imath \frac{7}{2}\pi}=\cdots
\end{eqnarray}

\noindent and

\begin{eqnarray}
f[-h]=e^{+\imath \frac{1}{2}\pi}=e^{-\imath \frac{3}{2}\pi}
=e^{+\imath \frac{5}{2}\pi}=e^{-\imath \frac{7}{2}\pi}=\cdots.
\end{eqnarray}

\noindent As we can see, in each class, the particles with different 
values of statistics parameter ($\nu=2s$) 
have unambiguously distinct braiding properties to obey 
the fractal distribution function determined by the parameter $h$. 

The phase of the wave function ( consider two-particle system ) 
changes by $+\pi\nu$ ( anti-clockwise exchange ) and  $-\pi\nu$  
( clockwise exchange ) in response to which way we braid in interchanging 
$x$ and $y$. On the one hand, the violation of the discrete symmetries of parity 
and time reversal is verified for such fractons, i.e. 

\begin{eqnarray}
 e^{\pm\imath\pi\nu}\longrightarrow e^{\mp\imath\pi\nu}.
\end{eqnarray}

\noindent We have applied these ideas in some systems 
of the condensed matter such as fractional quantum Hall effect\cite{R1} 
and Luttinger liquids\cite{R4}. Besides this we have also considered  
some connection with conformal field theories\cite{R3} . Another direction 
of research is the connection between {\bf Number Theory and Physics} as 
explicited by our formulation\cite{R1,R3}. There exists other
possibilities of application, for example, we are considering 
the study of entanglement states for {\it fracton quantum computing}\cite{R5}.

In conclusion, we have considered in this Letter, 
the fractal von Neumann entropy associated with the 
fractal distribution function for some universal classes $h$ of fractons. The 
derivation of the Eqs.(5) and (6) obtained in Ref.\cite{R1} using the Eq.(20) 
suggested in Ref.\cite{R2} and adapted to our formulation is another 
result of this paper. 
We have also determined the fractal-deformed 
Heisenberg algebras for these classes of fractons 
which obey specific fractal distribution function. 
A next step is to consider 
the construction of the nonlocal operators for fractons 
taking into account an angle function as discussed in\cite{R6}.

\end{document}